\documentclass{iau}
\usepackage{graphicx}

\usepackage{subfig}

\title[JD 11.~~Rotation Measures in AGN jets seen by VLA] 
{Rotation Measures in AGN jets \\ seen by VLA at 21 cm to 6 mm}

\author[Evgeniya V. Kravchenko, William D. Cotton \& Yuri Y. Kovalev]   
{Evgeniya V. Kravchenko$^{1}$, William D. Cotton$^2$ \and Yuri Y. Kovalev$^{1}$}

\affiliation{$^1$Lebedev Physical Institute,
\\ Profsoyuznaya 84/32, 117997 Moscow, Russia \\ email: {\tt evk@asc.rssi.ru} \\[\affilskip]

$^2$National Radio Astronomy Observatory, \\ 520 Edgemont Road,
Charlottesville, 22903-2475 Virginia, USA}

\pubyear{2014}
\volume{xxx}  
\pagerange{119--126}
\jname{Extragalactic jets from every angle} 
\editors{F. Massaro, C. C. Cheung, E. Lopez \& A. Siemiginowska, eds.}
\begin{document}

\maketitle

\begin{abstract}
We present Faraday Rotation Measure (RM) properties of seven active galactic nuclei (AGN),
observed with the NRAO VLA at three epochs in 2012-2014. Data was taken at 1.4, 2.2, 5.0, 8.2, 
15.4, 22.4, 33.5 and 43.1 GHz quasi simultaneously in full polarization mode.
For the first time RMs were calculated in a range of wavelengths covering more than one order 
of magnitude: from 21 cm up to 6 mm. 
We measured RM for each source and showed a tendency to increase its value toward high 
frequencies according to the law $|RM|\sim{\nu}^{a}$ with
$a=$1.6$\pm$0.1. For 0710+439, we observed an increase over the frequency range 
of 4 orders of magnitude and measured one of the highest RM ever,
$(-89\pm1)\cdot10^3$rad/m$^2$.\\
Analysis of different epochs shows variations of the value and the sign of RM
on short and long time-scales. This may be caused by changing physical conditions in the 
compact regions of the AGN jets, e.g. strength of magnetic field, particle density and so on.
\end{abstract}

\firstsection 
\section{Introduction}

Due to presence of highly magnetized media in the close vicinity of, and in AGN jets 
themselves, polarized emission is 
the subject of Faraday effects (\cite[Faraday 1993]{Faraday1933}). 
One of them is Faraday rotation, which causes rotation of the plane of polarization of 
an electromagnetic wave. As a result, the intrinsic position angle
(EVPA), $\chi_0$, of the jet electric vector will be rotated on a factor, 
depending on the plasma properties and observed wavelength. In the simplest case,
when magnetized plasma isn't intermixed with the emission region and depolarization effects
are small, 
the EVPA depends linearly on $\lambda ^{2}$: 
${\chi}_{observed} = \chi_0 + RM\cdot \lambda ^{2}$, where the constant $RM$ is
the Rotation Measure:
\[
RM = {{e^3}\over{8{\pi}^2{\epsilon}_0m^2c^3}} \int n_e \bf{B_{\parallel}} dl.
\]

Thus, {\it RM} is proportional to the magnetic field component, parallel to the 
line of sight, $\bf{B_{\parallel}}$ and particle volume density $n_e$ along the path $d\bf{l}$.
The study of Faraday Rotation can be done only in multi-frequency observations.

Study of the polarized emission variations in AGN jets on short time scales
is of particularly interest because it provides information about the jet structure and points to
the locations of the regions where the emission originates and how it propagates to an observer.

Variability of the electric vector position angle in blazars on weekly, monthly and year-time
intervals was shown more than once (e.g. \cite[D'Arcangelo et al. 2007]{DAr2007},
 \cite[Jorstad et al. 2007]{Jorstad2007},
 \cite[Agudo et al. 2014]{Agudo 2014}).
So far, only few a works have been done (e.g. \cite[G\'{o}mez et al. 2011]{Gomez2011}) 
analyzing the region of the jet where this 
variability goes from and it
remains to be the subject of future studies.

The goal of this work is tp probe the physical conditions in AGN jets 
and their structure by studying the Rotation Measures in different sources, 
through frequency and time.

\begin{figure}[t]
\begin{center}
 \includegraphics[scale=0.6]{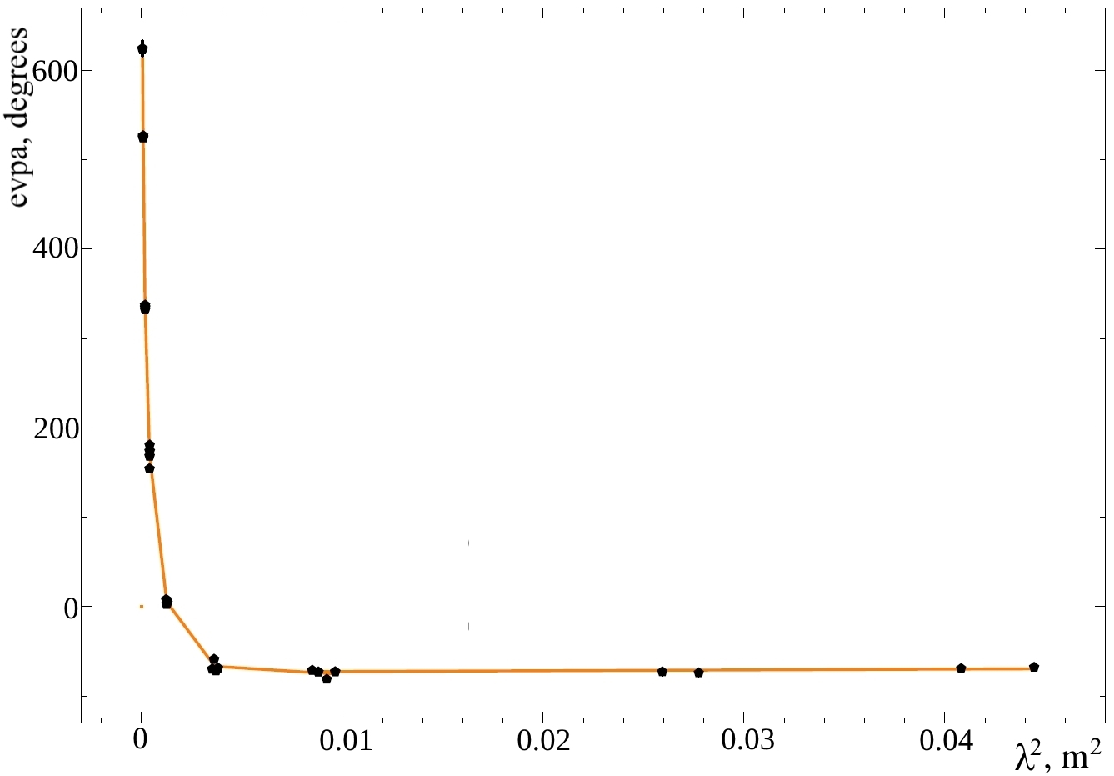}
 \includegraphics[scale=0.55]{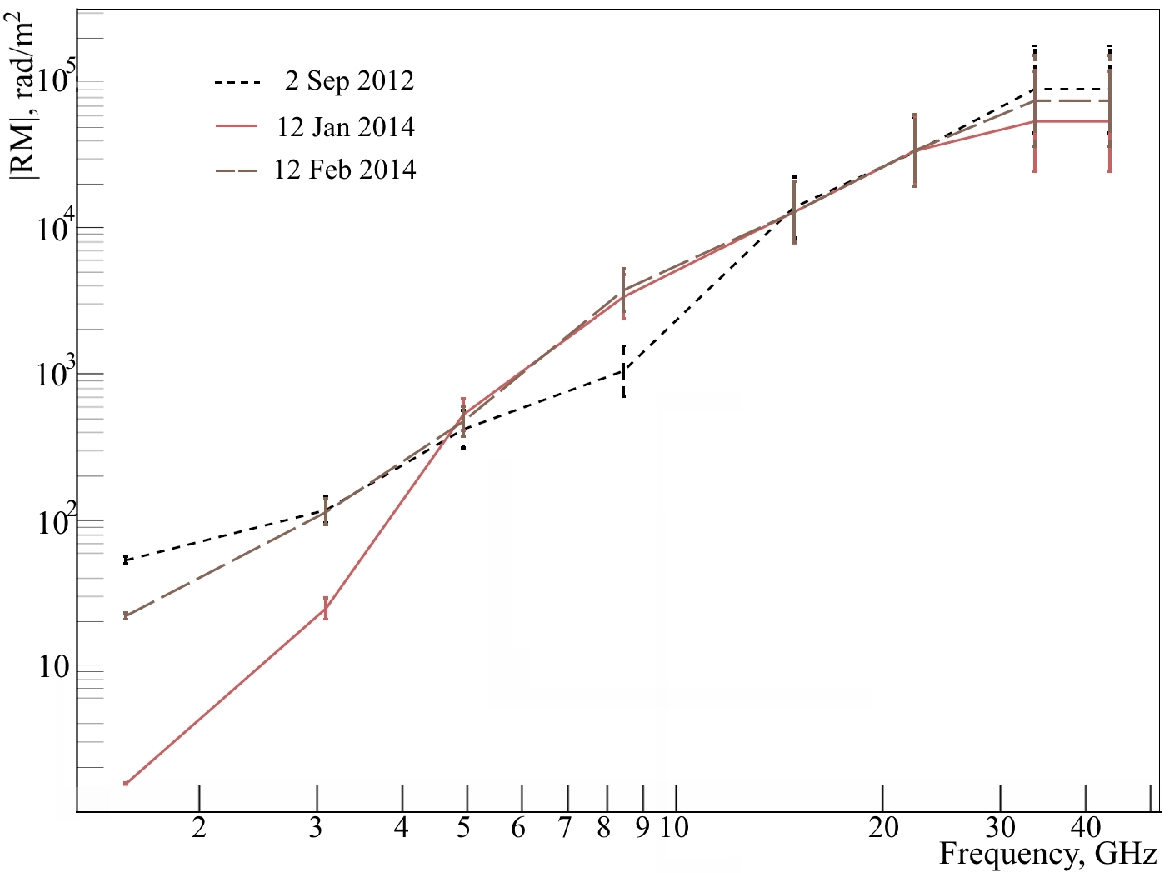} 
 \caption{(left) Example of reconstructed dependence of electric
 vector position angle with wavelength squared for 0710+439.
 (right) Derived Rotation Measure versus frequency for 0710+439 at three epochs in logarithmic scale.}
   \label{fig1}
\end{center}
\end{figure}

\section{Observations}

We use data from the EVLA polarization calibration program 
(\cite[Taylor \& Myers 2010]{TaylorMyers2010}), available from 
NRAO Archive\footnote{https://archive.nrao.edu/archive/advquery.jsp} under project TPOL0003.
Data consists of observations made at eight frequencies quasi
simultaneously, with switching between frequencies made within 30 minutes sequentially. 
The bandwidth in each frequency band is 256 MHz. To avoid bandwidth smearing 
(\cite[Gardner \& Whiteoak 1966]{GardnerWhiteoak1966}) 
we split bands onto 2 to 4 frequency channels, depending on the wavelength. The resulting
frequency setup is given in Table \ref{tab1}.

Target sources are presented in Table \ref{tab2}.
There are 13 epochs of observations available with the frequency setup given above,
starting in 2011. The configuration of the VLA during these observations
goes through all possible setups. 
Here we present results from last 3 epochs only:
2 September 2012, 12 January 2014 and 12 February 2014. 

\begin{table}
  \begin{center}
  \caption{Frequency setup.}
  \label{tab1}
 {\scriptsize
  \begin{tabular}{|c|c|c|}\hline 
{\bf Band} & {\bf Central frequency} & {\bf Channel} \\ 
 & {\bf of channels, GHz} & {\bf bandwidth, MHz} \\ \hline
 L & 1.423, 1.485, 1.801, 1.863 & 64 \\\hline
 S & 3.055, 3.124, 3.197, 3.253 & 64 \\\hline
 C & 4.863, 4.925, 4.991, 5.053 & 64 \\\hline
 X & 8.365, 8.431, 8.497, 8.559 & 64 \\\hline
 K$_u$ & 14.863, 14.927, 14.991, 15.055 & 64 \\\hline
 K & 22.373, 22.457, 22.543 & 85 \\\hline
 K$_a$ & 33.473, 33.559, 33.645 & 85 \\\hline
 Q & 43.215, 43.343 & 128 \\  \hline
  \end{tabular}
  }
 \end{center}
\end{table}

\begin{table}
  \begin{center}
  \caption{Target sources.}
  \label{tab2}
 {\scriptsize
  \begin{tabular}{|c|c|c|c|}\hline 
{\bf B1950 source name} & {\bf Alias} & {\bf Optical class} & {\bf Spectral type}\\ \hline
0552$+$398 &   DA 193   & Quasar & peaked \\\hline
0710$+$439 & B3 0710+438 & Radio galaxy & steep\\\hline
0736$+$017 &   OI 061   & Quasar & double-peaked \\\hline
0851$+$202 &   OJ 287   & BL Lac & flat \\\hline
0923$+$392 &  4C +39.25 & Quasar & flat one-peaked\\\hline
1253$-$055 &   3C 279   & Quasar & double-peaked \\\hline
1308$+$326 &   OP 313  & Quasar & flat \\ \hline
  \end{tabular}
  }
 \end{center}
\end{table}

\section{Data reduction}

Data calibration and imaging is done using Obit 
package\footnote{http://www.cv.nrao.edu/~bcotton/Obit.html}
 (\cite[Cotton 2008]{Cotton2008}).


Since our observations cover wide frequency range, we probe regions in AGN jets
with different Faraday depths and different physical properties. 
Supposing the external to the jet nature of Faraday media, EVPA 
should depend from wavelength-squared linearly.
To sum up these two assumptions, EVPA is linear with ${\lambda}^2$ within different intervals, but
may arbitrarily change through the whole range of ${\lambda}^2$.
Thus we identify the Rotation Measure as a linear slope in the dependence EVPA-${\lambda}^2$ at
individual ${\lambda}^2$ intervals. We let EVPA 
wrap on 180$^{\circ}$ between different channels and pick the
solutions with the minimal ${\chi}^2$ and minimal number of wraps. 
Example of the reconstructed curve is shown on Figure \ref {fig1}.

Length and number of the individual intervals, where RM were obtained,
varies from source to source. For instance for 0710$+$439, given on Figure \ref {fig1},
RM were identified at seven ${\lambda}^2$ intervals, solutions for which 
with frequency are shown on Figure \ref {fig1} with solid curve.

Target sources are roughly point-like even for the most extended VLA configuration. 
Thus, we carry out the analysis only on the central region of the map, which is composed of 
the optically thin and thick components of the parsec-scale jet. To conclude what component dominates
in the map we analysed fractional polarization and spectral index of the sources.

Spectra have different types: flat or steep behavior, one or double
peaked and are given in Table \ref{tab2}.
The distribution of the fractional polariation, given on Figure \ref{fig2}, also indicates that we 
observed a mix of regions with different optical depths. 

During the analysis, each epoch were considered independently. 
The typical value of rotation from Galactic media (\cite[Taylor et al. 2009]
{Tayloret2009}) is a few radians per meters squared which is
significant only at low frequencies. 
We didn't correct our measurements for it.

\section{Results}

\begin{figure}[t]
\begin{center}
 \includegraphics[scale=0.63]{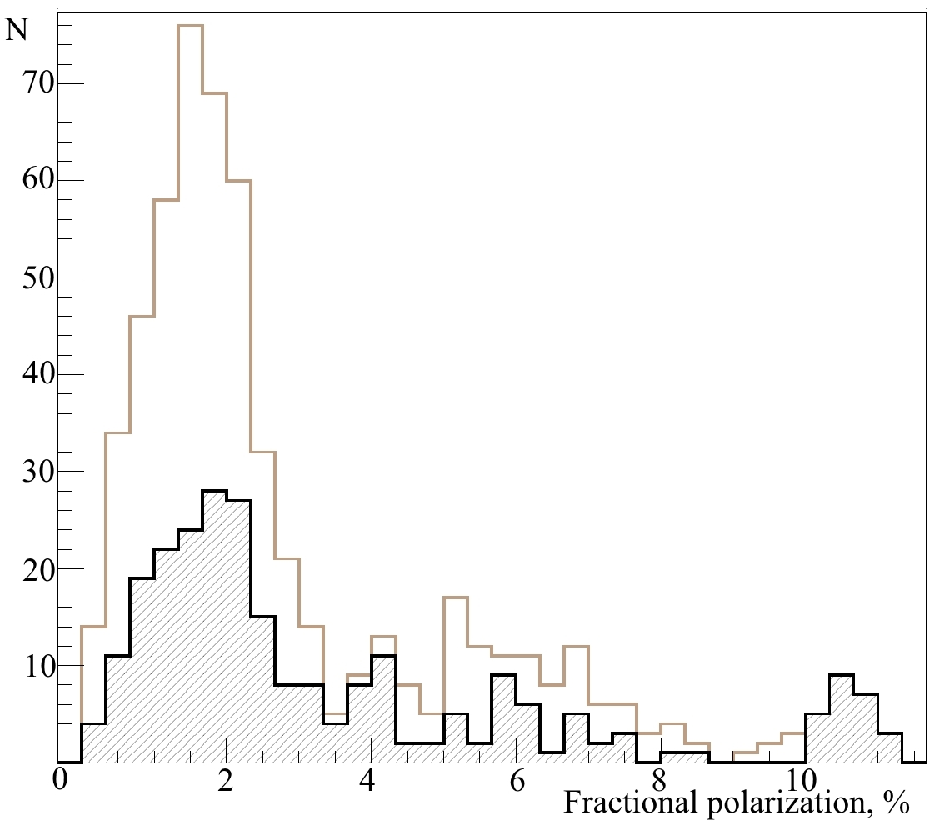} 
 \includegraphics[scale=0.58]{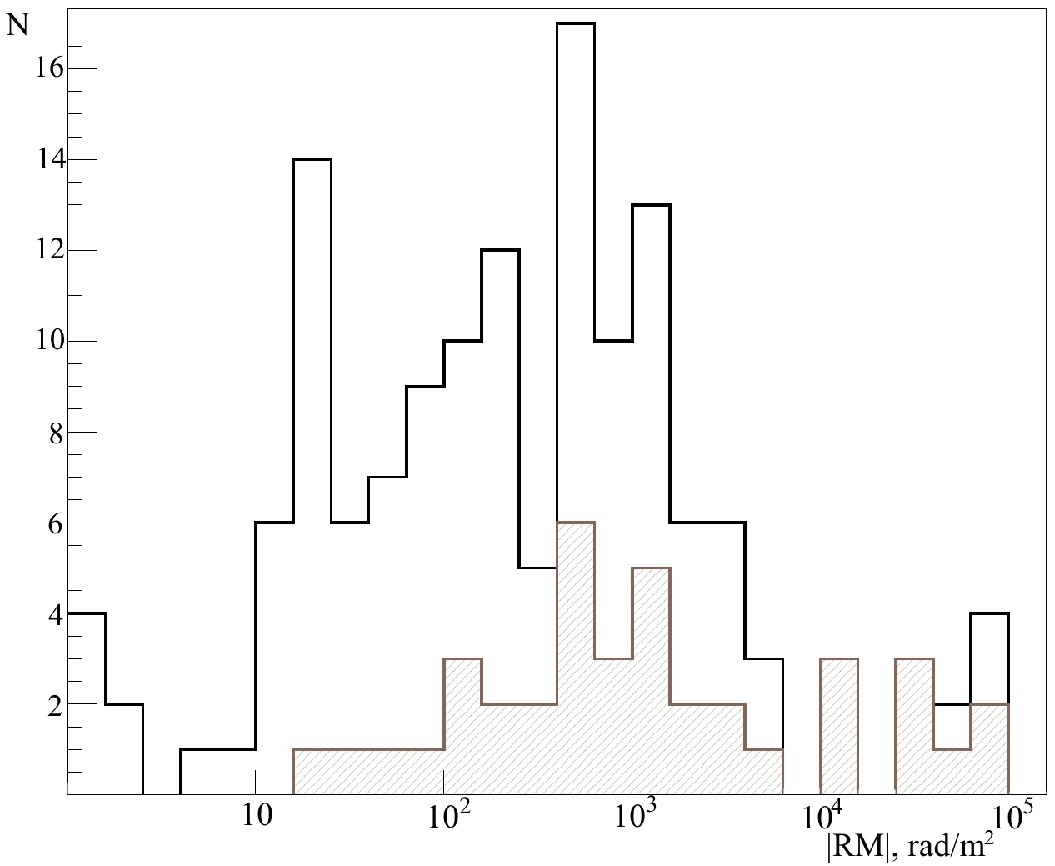} 
 \caption{(left) Distribution of the fractional polarization for all sources at eight frequency bands
 and at three epochs of obsevations. Hatched region indicates 14 - 43 GHz frequency range,
 white - 1.4 - 14 GHz range.
 (right) Distribution of the obtained Rotation Measures for all target sources at
 eight frequencies at three epochs. Shaded region indicates measurements in 14 - 43 GHz range,
 while white region represents 1.4 - 14 GHz frequency range.}
   \label{fig2}
\end{center}
\end{figure}

We have estimated RMs for all target sources through the whole ${\lambda}^2$-interval at 
three epochs, the distribution of which is represented on Figure \ref{fig2}. 
Variations of a few orders of magnitude in the value of RM can be seen
there, meaning strong variations of  
physical conditions in AGN jets among sources.

For the majority of the cases, wraps of 180$^{\circ}$ in EVPA at some frequency bands
relative to the other bands reduces ${\chi}^2$ considerably.
Moreover, an analysis of different epochs shows good agreement between estimated values of RM
and it's trend with frequency.
It is important to note that sometimes we see non-trivial EVPA-${\lambda}^2$ behavior 
which can not be fit by a linear slope. This means that other Faraday effects take place there,
such as mixing of the rotating media with the emission region, multi component Faraday media with different
optical depths and others.

A tendency of RM to increase in value with frequency $|RM|\sim{\nu}^{a}$,
example of which is given on Figure \ref{fig1}, is found for all sources.
We have used RM values for all available ${\lambda}^2$ intervals to determine
power $\it{a}$ for every target performing a linear regression of lg$|RM|$-lg$(\nu)$ curves.
An average over all targets and three epochs is found to be 1.6$\pm$0.1.

Because of synchrotron self absorption increasing towards the jet base, we see regions  
being located closer to the central engine at higher frequencies. 
A possible explanation of 
the trend of higher RM at higher frequency is that higher frequencies
probe regions closer to central engine with a denser media and
stronger magnetic fields.  
This trend indicates variations of $n_e \bf{B_{\parallel}}$$d$$\bf{l}$
by up to five orders of magnitude with distance from the central black hole.
Theoretical estimations give a value of $\it{a}=2$ (\cite[Jorstad et al. 2007]{Jorstad2007}), 
asuming an outflowing sheath around a conically or spherically expanding jet with a helically-shaped
magnetic field. Our low value of $\it{a}$ may result from underestimated contribution of the emission
from optically thin regions. We plan to conduct a detailed study of this effect in the future.

{\underline{\it Case of 0710$+$439}}. For this source we obtained the
following values of RM (in rad/m$^2$) in the  33 - 43 GHz range:
$(-89\pm1)\cdot10^3$, $(-54\pm3)\cdot10^3$ and $(-74\pm1)\cdot10^3$ at three 
consecutive epochs given above.
These values were measured independently and are in good agreement.
The behavior of RM over the full frequency range 
for 0710$+$439 is shown on Figure \ref {fig1}. 

So far, high values of Rotation Measure in AGNs were observed in 
only a few sources (e.g. \cite[Trippe et al. 2012]{Trippe2012}, 
\cite[Jorstad et al. 2007]{Jorstad2007}, \cite[Attridge et al. 2005]{Attridge2005}) 
with the highest value of $5.6\cdot10^{5}$ rad/m$^2$
for Sagittarius $A^{\star}$ (\cite[Marrone et al. 2006]{Marrone2006}). 

Observations of all these sources, except at the center of our Galaxy, are non-simultaneous
with the differences between the observational epochs at different
frequencies of up to a few months and even years. 
Our observations have an advantage over these measures, and thereby don't include time variability 
of the sources which may result in dramatically different interpreted and real 
polarization pictures. Thus, 0710$+$439 has one of the highest recorded values of RMs so far 
made under assumption that Faraday screen doesn't mix with the jet emission region. 
It may mean that 0710$+$439 holds dense media with strong magnetic
fields in the innermost jet regions.\\ 

To continue this study we plan to analyze the other 10 epochs of
observations for a more detailed picture 
of Rotation Measure behavior in time.

This research was supported by Russian Foundation for Basic Research, research project No. 
14-02-31789 mol\textunderscore a. The VLA is an instrument 
of the National Radio Astronomy Observatory, a facility of the National Science Foundation operated 
under cooperative agreement by Associated Universities, Inc.

\end{document}